\documentclass[twocolumn]{revtex4}

\usepackage{graphicx}

\begin{document}

\title{Viability of complex self-interacting scalar field as dark matter.}
\author{F. Briscese }\email{fabio@fisica.unam.mx}
\affiliation{Instituto de F\'{\i}sica, Universidad Nacional
Autonoma de Mexico, Apdo. Postal 20-364, 01000 M\'exico D.F.,
Mexico.
\\
and
\\
Istituto Nazionale di Alta Matematica Francesco Severi \\
Gruppo Nazionale di Fisica Matematica,\\
citt$\grave{a}$ Universitaria, c.a.p. 00185, Rome, Italy.}


\begin{abstract}
\begin{center}
\end{center}
We study the viability of a complex scalar field $\chi$ with
self-interacting potential $ V = m^\chi_0/2 \, |\chi|^2 + h \,
|\chi|^4$ as dark matter. The scalar field is produced at
reheating through the decay of the inflaton field and
 then, due to the self-interaction, a Bose-Einstein condensate of
$\chi$ particles forms. The condensate represents dark matter in
that model. We analyze the cosmological evolution of the model,
stressing how, due to the presence of the self-interaction, the
model naturally admits dark matter domination at late times, thus
avoiding any fine tuning on the energy density of the scalar field
at early times. Finally we give a lower bound for the size of dark
matter halos at present time and we show that our model is
compatible with dark matter halos greater than $0.1 \, Kpc$ and
with BBN and CMB bounds on the effective number of extra neutrinos
$\Delta_\nu^{eff}$. Therefore, the model is viable and for $h
\simeq 10^{-4}-10^{-12}$ one obtains a mass $m^\chi \simeq
m^{\chi}_0 \simeq 1-10^{-2} \, eV$ for dark matter particles from
radiation-matter equality epoch to present time, but at
temperatures $T_\gamma \gg 10 \, eV$, where $T_\gamma$ is the
photons temperature, thermal corrections to $m^\chi_0$ due to the
self-coupling $h$ are dominant.

\end{abstract}

\maketitle

\section{Introduction}

Dark matter is one of the most fundamental ingredients of modern
cosmology. Evidence for its existence comes from cosmological and
astrophysical observations, e.g. cosmic microwave background (CMB)
temperature anisotropy \cite{cmb}, large scale structures of the
universe \cite{sdss} and measurements of galaxy rotation curves
\cite{rotationv}. There are many models that aim to explain the
nature of dark matter, e.g. weakly interacting massive particles
(WIMPS), axions \cite{bertone} and modified versions of general
relativity \cite{odisntsov}. A valid alternative that has been
widely studied, is to consider a scalar field as dark matter
candidate \cite{scalar} and recently it has been studied the case
in which a  non-self-interacting scalar field forms a
Bose-Einstein condensate \cite{SDFMcondensate}. Scalar field dark
matter (SFDM) models are competitive with $\Lambda CDM$ model to
explain observational evidence of dark matter at cosmological
level, up to linear perturbations.

Here we examine the possibility of describing dark matter through
a complex self-interacting scalar field $\chi$ that forms a
Bose-Einstein condensate at early times just after reheating. The
scalar field has a renormalizable self-interacting potential
$v(\chi,\bar\chi) = m^{\chi 2}_0 \, |\chi|^2 /2 +  h \, |\chi|^4$
and, as we will discuss extensively, the presence of the
self-interaction has many important consequences for the model.
The first consequence it that  it allows the formation of a $\chi$
particle condensate at early times. As was first studied in
ref.\cite{Peloso}, if the $\chi$ field is coupled with the
inflaton field it is possible that at reheating the $\chi$ field
is generated with a charge asymmetry $Q^\chi \equiv n^\chi -
n^{\bar\chi}$, where $n^\chi$ and $n^{\bar\chi}$ are the number
density of $\chi$ and $\bar\chi$ particles. If $Q^\chi$ is larger
than some critical value, then a $\chi$-particle condensate forms
just after reheating and the scalar field configuration will be
that of a condensate in equilibrium with a  thermalized gas of
$\chi$ and $\bar\chi$ particles at temperature $T_\chi$.

A second key consequence of the self-interaction $h \, |\chi|^4$
is that it gives  thermal corrections to $m^\chi_0$ that become
important at high temperatures, giving an effective $\chi$ mass
$m^\chi \sim T_\chi$ and therefore a condensate energy density
$\rho^\chi_c \sim T_\chi^4$ scaling like radiation. Therefore, at
high temperatures $T_\chi \gg T_{1\chi} \gg m_o^\chi$, where
$T_{1\chi}$ depends on the model parameters, both the condensate
energy density $\rho^\chi_c$ and that of the thermalized $\chi$
and $\bar\chi$ particles $\rho^\chi_{th}$ scale like radiation
$\rho^\chi_c \sim \rho^\chi_{th} \sim T_\chi^4$. This implies
that, supposing that at early times the $\chi$ field is
subdominant, the condensate cannot dominate over radiation as long
as $T_\chi > T_{1\chi}$. We will show that one can fix the
parameters of the model in such a way that the condensate gives
the expected energy density  $\rho^\chi_c = \rho^{DM} \simeq 0.323
\, eV^4$ at radiation-matter equality time without any fine tuning
on the condensate energy density at early times. This gives a
clear interpretation of the late time dominance of dark matter:
the condensate dominates at late times since as long as $T_\chi >
T_{1\chi}$ its energy density $\rho^\chi_c$ scales like radiation.

Finally we determine a lower limit $L_H$ for the size of dark
matter halos at present epoch and we study the dependence of
$\Delta_\nu^\chi$ on $L_H$, where $\Delta_\nu^\chi$ is the
contribution of the scalar field $\chi$ to the effective number of
extra neutrinos. We show that $\Delta_\nu^\chi \simeq 3.34 \,
\left( L_H/Mpc  \right)^{2/3}$ and does not depend on the coupling
$h$, therefore it is possible to lower $\Delta_\nu^\chi$ below the
big bang nucleosynthesis (BBN) bounds just diminishing $L_H$.
Since $L_H$ is a lower limit for the dark matter halo sizes, any
value $L_H < 100 Kpc$ is acceptable and therefore any $L_H \leq
0.1 \, Kpc$ gives a $\Delta_\nu^\chi$ within   BBN bounds.

Choosing $L_H \simeq 0.1 \, Kpc$, at dark matter domination the
$\chi$ mass variates in the range $m^\chi \simeq 1-10^{-2} \, eV$
for $h \simeq 10^{-4}-10^{-12}$. This mass should be compared with
the case of scalar field models with no self-interactions, for
which one has an extremely low mass of about $10^{-23} \, eV$
\cite{SDFMcondensate}. Moreover the transition of the $\chi$
condensate from radiation $\rho^\chi_c \sim T_\chi^4$ to matter
$\rho^\chi_c \sim T_\chi^3$ occurs at temperatures $T_\gamma
\simeq 10 \, eV$, where $T_\gamma$ is the photons temperature,
namely just before radiation matter equality. This makes it
possible to obtain the expected value of  $\rho^\chi_c = \rho_{DM}
= 0.232 \, eV^4$ at equality epoch without any fine tuning.

This Letter is organized as follows: in section
\ref{sectioncondensate} we describe the physics of a system
composed of the  Bose-Einstein $\chi$-particle condensate in
equilibrium with $\chi$ and $\bar\chi$ thermalized particles. In
section \ref{cosmologicalevolution} we  describe the cosmological
evolution of the condensate. In section
\ref{sectioncondensateformation} we  discuss the conditions under
which the condensate forms, assuming that $\chi$ particles are
produced at reheating via inflaton decay. In section
\ref{sectionconextrarelativisticdegreesof} we derive the
contribution of the $\chi$ condensate and of the thermalized gas
of $\chi$ and $\bar\chi$ particles to the effective number of
extra neutrinos $\Delta_\nu^{eff}$. In section \ref{dmhalos} we
determine the lower bound $L_H$ for dark matter halos at present
times and in section \ref{sectionrealisticmodel} we present a
choice of the model parameters that gives a realistic model.
Finally in section \ref{sectionconclusions} we conclude.

\section{Bose-Einstein Condensate}\label{sectioncondensate}

Consider a scalar field with Lagrangian

\begin{equation}\label{lagrangian}
L = \frac{1}{2} \partial_\mu \chi \partial^\mu \chi - \frac{1}{2}
m_0^\chi \, |\chi|^2 - h \, |\chi|^4
\end{equation}

with $h \ll 1$. We assume that the $\chi$ particles are weakly
self-interacting and that their mass varies adiabatically. One can
then define the phase space distributions $f_\chi$ and
$f_{\bar{\chi}}$ of the $\chi$ particles and ${\bar{\chi}}$
antiparticles so that the energy, number and charge density of the
complex $\chi$ field are respectively

\begin{equation}\label{energydensity}
\begin{array}{ll}
\rho_\chi = (2\pi)^{-3} \, \int d^3p \, E_\chi(p) \,
\left[f_\chi(p) + f_{\bar{\chi}}(p)\right]\\
\\
n_\chi = (2\pi)^{-3} \, \int d^3p  \, \left[f_\chi(p) +
f_{\bar{\chi}}(p)\right]\\
\\
Q^\chi = (2\pi)^{-3} \, \int d^3p \, \left[f_\chi(p) -
f_{\bar{\chi}}(p)\right]
\end{array}
\end{equation}

The phase space distributions for a $\chi$-particle condensate in
equilibrium with thermalized $\chi$ and  $\bar\chi$ particles with
temperature $T_\chi$ are

\begin{equation}\label{fchitot}
\begin{array}{ll}
f_\chi(p) = f_\chi^{BE}(p) + (2\pi)^3 \, Q_c \, \,\delta^3(p)\\
\\
f_{\bar{\chi}}(p) = f_{\bar{\chi}}^{BE}(p)
\end{array}
\end{equation}

where $f_\chi^{BE}(p) = 1/\left[e^{\beta\,(E-\mu)}-1 \right]$ and
$f_{\bar{\chi}}^{BE}(p) = 1/\left[e^{\beta\,(E+\mu)}-1 \right]$,
$\beta = 1/T_\chi$, $\mu$ is the chemical potential and $Q_c$ is
the number density of the $\chi$ particles of the condensate.
Following ref.\cite{Peloso} one can calculate the thermal
correction to the $\chi$ mass as $\left(m^{\chi}_{th}\right)^2
\simeq 4 h \, \int \frac{d^3p}{(2\pi)^3 \, 2 E} \left( f_\chi(p) +
f_{\bar{\chi}}(p) \right) \simeq \left( 2 h Q_c + \frac{1}{3}h
T_\chi^3\right)/m^\chi_{th}$, that gives

\begin{equation}\label{mthermal}
\begin{array}{ll}
m^\chi_{th}  \simeq \alpha \, T_\chi,\qquad \alpha \equiv \left[
\, h\, \left(2 \frac{Q_c}{T^3_\chi} + \frac{1}{3}
\right)\right]^{1/3}
\end{array}
\end{equation}

Therefore the effective mass $m^\chi$ of the $\chi$ and $\bar\chi$
particles will be

\begin{equation}\label{effectivemass}
\begin{array}{ll}
m^\chi \simeq m_0^\chi \qquad \qquad \qquad\quad for \qquad T_\chi
\leq
\frac{m_0^\chi}{\alpha}\\
\\
m^\chi \simeq m_{th}^\chi(Q_c,T_\chi,h) \qquad for \qquad T_\chi
\gg \frac{m_0^\chi}{\alpha}
\end{array}
\end{equation}

The number density of $\chi$ particles is

\begin{equation}\label{thermalnumberdensity}
\begin{array}{ll}
n^\chi = Q_c + n^\chi_{th},\qquad n^\chi_{th} \equiv (2\pi)^{-3}
\int d^3p \, f^{BE}_\chi(p) ,
\end{array}
\end{equation}

the number density of $\bar\chi$ particles is

\begin{equation}\label{thermalnumberdensity2}
\begin{array}{ll}
n^{\bar\chi} = n^{\bar\chi}_{th} \equiv (2\pi)^{-3} \, \int d^3p
\, f^{BE}_{\bar\chi}(p) ,
\end{array}
\end{equation}

while the energy density of the $\chi$ field is

\begin{equation}\label{thermalenergydensity}
\begin{array}{ll}
\rho^\chi = m^\chi Q_c + \rho^\chi_{th}\\
\\
\rho^\chi_{th}= (2\pi)^{-3} \, \int d^3p \, E_\chi(p) \,
\left[f^{BE}_\chi(p) + f^{BE}_{\bar{\chi}}(p) \right] .
\end{array}
\end{equation}

Also  the charge density is

\begin{equation}\label{thermalchargedensity}
\begin{array}{ll}
Q^\chi = Q_c + Q^\chi_{th}\\
\\
Q^\chi_{th}= n^\chi_{th} - n_{th}^{\bar{\chi}} = (2\pi)^{-3} \int
d^3p \, \left[f^{BE}_\chi(p) - f^{BE}_{\bar{\chi}}(p) \right] .
\end{array}
\end{equation}

Note that, since the thermal corrections to $m_0^\chi$ depend on
$Q_c$, $T_\chi$ and $h$, then $f^{BE}_\chi(p)$ and
$f^{BE}_{\bar{\chi}}(p)$ will depend on $Q_c$ and $h$ via the
effective mass $m^\chi(Q_c,T_\chi,h)$. Therefore $n^{\chi}_{th}$,
$\rho^{\chi}_{th}$ and $Q^{\chi}_{th}$ in general also depend on
$Q_c$ and $h$. In any case, at temperatures $T_\chi \gg m^\chi
\geq \mu > 0$ one can neglect both $m^\chi$ and $\mu$ and recover
the usual results \cite{dodelson}


\begin{equation}\label{thermalrelativisic}
n^\chi_{th} = n^{\bar\chi}_{th} = \frac{\zeta(3)}{\pi^2} \,
T_\chi^3, \quad \rho^\chi_{th} = \frac{\pi^2}{15} \, T_\chi^4,
\quad Q^\chi_{th} = \frac{\mu(T_\chi)}{3} \, T_\chi^2
\end{equation}

For simplicity we also define the condensate contribution to the
number, charge and energy density as

\begin{equation}
n_\chi^c \equiv Q_{\chi}^c  \equiv Q_c, \qquad \rho^\chi_{c}
\equiv m_\chi Q_c
\end{equation}

\section{Cosmological evolution}\label{cosmologicalevolution}

In this section we resume the main features of the cosmological
evolution of the scalar field $\chi$. The phase space
distributions given in Eq.(\ref{fchitot}) are solutions of the
relativistic Boltzmann equations in FRW metric for $T_\chi \sim
1/a(t)$, $Q_c \sim T_\chi^3 \sim 1/a(t)^{3}$ and $\mu = 0$, at any
temperature $T_\chi \gg m^\chi \geq \mu > 0$ and for any initial
value of $Q_c/T^3_\chi$. Note also that $Q_c/T^3_\chi$ remains
constant as long as $T_\chi \gg m^\chi$, therefore from
Eq.(\ref{mthermal}) it is evident that $\alpha$ is constant in the
same range of temperatures. In what follows we assume that the
coupling $h$ is small enough to give $\alpha \ll 1$. Therefore
from Eq.(\ref{mthermal}) it follows that $T_\chi \gg m^\chi_{th}$
at any time. We define the parameter

\begin{equation}\label{kdefinition}
k \equiv T_\chi/T_\gamma
\end{equation}

where $T_\gamma$ is the photons temperature, and we note that $k$
is also constant for $T_\chi \gg m^\chi_0$. Moreover we define the
following temperatures

\begin{equation}\label{Tdefinition}
\begin{array}{ll}
T_{1\chi} \equiv m_0^\chi/\alpha, \quad T_{2\chi} \equiv
m_0^\chi\\
\\
T_{1\gamma} \equiv T_{1\chi}/k, \quad T_{2\gamma} \equiv
T_{2\chi}/k
\end{array}
\end{equation}

with $T_{1\chi} \gg T_{2\chi}$ and $T_{1\gamma} \gg T_{2\gamma}$
since $\alpha \ll1$. Finally we define $t_1$ as the time when
$T_\chi = T_{1\chi}$ and $T_\gamma = T_{1\gamma}$ and $t_2$ as the
time when $T_\chi = T_{2\chi}$ and $T_\gamma = T_{2\gamma}$.

Armed with these definitions we  describe the cosmological
evolution of the scalar field $\chi$. There are three
important epochs in which the $\chi$ field behave differently.\\
--- At early times $t \ll t_1$, when  $T_\chi \gg T_{1\chi}$ the
$\chi$ mass is dominated by thermal corrections so $m^\chi \simeq
m_{th}^\chi = \alpha T_\chi \ll T_\chi$. That implies that the
energy density of the condensate evolves as radiation since
$\rho^\chi_c \simeq m_{th}^\chi \, Q_c \sim T_\chi^4$. Of course,
since $T_\chi \gg m^\chi$, Eqs.(\ref{thermalrelativisic}) are
valid and  $\rho_{th}^\chi \sim T_\chi^4$. In conclusion the whole
$\chi$ field evolves as radiation with $\rho^\chi \sim
T^{4}_\chi$.\\
--- At temperatures $T_{1\chi} \gg T_\chi \gg T_{2\chi}$ one has
$m_0^\chi \gg m_{th}^\chi$, therefore the mass of the $\chi$
particles is simply $m_0^\chi$. Moreover one still have $T_\chi
\gg m_\chi$, therefore Eqs.(\ref{thermalrelativisic}) are still
valid. Then in this case $\rho^{\chi}_{th} \sim T_\chi^4$ still
evolves as radiation but $\rho^\chi_c \simeq m^\chi_0 \, Q_c \sim
T_\chi^3$ evolves as matter. Therefore at the temperature
$T_{1\chi}$ the condensate passes from a radiation-like to a
matter-like evolution.\\
--- At temperatures $T_\chi < T_{2\chi}$ below $m_0^\chi$,
also the thermalized $\chi$ and $\bar\chi$ particles begins to
evolve as matter with $\rho_{th}^\chi \sim a^{-3}$
. Therefore the temperature $T_{2\chi}$ characterizes the
transition of $\rho_{th}^\chi$ from radiation-like to matter-like
fluid. In this range of temperatures the total energy density of
the scalar field evolves as matter, i.e. $\rho^\chi \sim a^{-3}$.
In particular, if at $T_\chi \simeq T_{2\chi}$ one has
$\rho^{\chi}_c \gg \rho^{\chi}_{th}$,  one can take $\rho^\chi
\simeq \rho^{\chi}_c(t_2) (a(t_2)/a(t))^3$ for any time $t > t_2$.

To summarize, the thermalized gas of $\chi$ particles and
$\bar{\chi}$ antiparticles becomes non-relativistic at
temperatures below $m_0^{\chi}$ as usual, but the condensate still
evolves as matter at temperatures  $T_{1 \chi} \gg T_\chi \gg
T_{2\chi}$ well above $m_0^\chi$. This last feature is typical of
this model and it is due to the fact that thermal corrections to
the mass are important only at very high temperatures, i.e. above
$T_{1\chi}$. Of course the thermal corrections to $m^\chi_0$ are
due to the presence of the $h \, |\chi|^4$ self-coupling. If
self-interactions are turned off, there are no thermal corrections
to $m_0^\chi$, therefore  the condensate always evolves as matter
and this implies a severe fine tuning on its energy density at
early times. Moreover the self-interaction is important for a
second reason. Since the condensate is formed right after
reheating, one should explain why it starts to dominate just at
radiation-matter equality. This question is easily answered in
that context. In fact, because of the self-interaction, the
condensate evolves as a relativistic fluid at high temperatures
and it cannot dominate over radiation before $t_1$, i.e. at
temperatures $T_\chi
> T_{1\chi}$ (or $T_\gamma > T_{1\gamma}\equiv T_{1\chi}/k$).
Therefore one can choose the coupling constant $h$ and $m^\chi_0$
properly, in order to ensure a dark matter domination at
temperatures $T_\gamma \simeq 0.698 \, eV$. This helps to explain
the cosmological coincidence problem without any fine tuning on
$\rho^\chi$ at early times.

\section{Condensate formation}\label{sectioncondensateformation}

In the model  that we are presenting we suppose that the scalar
field $\chi$ is produced at reheating via the inflaton decay. The
$\chi$ and $\bar\chi$ particles are produced with a charge
asymmetry  $Q^\chi >0$ via an Affleck-Dine mechanism
\cite{affleck}, and then, due to self-interactions, they forms a
$\chi$-particle condensate. The conditions under which the
condensate is formed are studied in ref.\cite{Peloso}.  Since the
charge and energy densities are conserved, the quantity $R \equiv
Q^\chi/\rho_\chi^{3/4}$ remains constant as long as $T_\chi \gg
m^\chi$. In ref.\cite{Peloso} it is found that the condensate
forms if the $\chi$ field is produced at reheating with

\begin{equation}\label{Rcondition}
R \geq 0.2 \, h^{1/2} .
\end{equation}
It is also found  that, if $R \geq 1/2$ one has $Q^\chi \gg
n^\chi_{th}$, i.e., the majority of the $\chi$ particles are in
the condensate. After the condensate formation, the phase space
distributions of the $\chi$ and $\bar{\chi}$ particles are given
by Eq.(\ref{fchitot}) and Eq.(\ref{Rcondition}) reads

\begin{equation}\label{Rcondition2}
R \equiv \frac{Q^\chi}{\rho_\chi^{3/4}} = \frac{Q_c/T_\chi^3 +
\mu(T_\chi)/3 T_\chi}{\left(
 \alpha \, Q_c/T_\chi^3 + \pi^2/15  \right)^{3/4}} > 0.2
 \, h^{1/2}
\end{equation}
Therefore, any realistic choice of the model parameters should
fulfill Eq.(\ref{Rcondition2}) for any $T_\chi \gg T_{1\chi}$. We
stress that the presence of the self-interaction is fundamental in
this model for the $\chi$-particle condensate formation.

\section{Effective number of extra neutrinos}\label{sectionconextrarelativisticdegreesof}

In the range $T_\chi > T_{2\chi} \equiv m_0^\chi$, the energy
density of thermalized $\chi$ particles evolves as radiation and
therefore the contribution of $\rho^\chi_{th}$  to the effective
number of extra neutrino $\Delta_\nu^{eff}$ is

\begin{equation}\label{extrarelativisticNthermal}
\Delta_\nu^{th} = \frac{16}{7} \left(\frac{T_\chi}{T_\nu}\right)^4
.
\end{equation}

In the range of temperatures  $T_\chi \gg T_{1\chi} \equiv
m_0^\chi/\alpha$, when $T_\chi \gg m^\chi_{th} \equiv \alpha
T_\chi \gg m_0^\chi$, also the $\chi$ condensate evolves as
radiation and one can write its contribution to $\Delta_\nu^{eff}$
as

\begin{equation}\label{extrarelativisticNcondensate}
\Delta_\nu^{c} = \frac{240}{7\pi^2}
\left(\frac{T_\chi}{T_\nu}\right)^4 \, \frac{\alpha \,
Q_c}{T_\chi^3} .
\end{equation}

Of course, the $\chi$ field does not contribute at all to
$\Delta_\nu^{eff}$ for $T_\chi < m_0^\chi$. Therefore,
Eqs.(\ref{extrarelativisticNthermal}) and
(\ref{extrarelativisticNcondensate}) can be used to constrain the
model with cosmological data. For example at BBN  one should
impose the condition $\Delta_\nu^{eff} = 0.054^{+1.4}_{-1.2}$
\cite{BBN} and at decoupling $\Delta_\nu^{eff}$ should be
constrained with the less stringent CMB value $\Delta_\nu^{eff} =
1.30^{+0.86}_{-0.88}$ \cite{cmb}.

\section{Dark Matter Halos}\label{dmhalos}

Following the analysis given in \cite{Arbey} we can study the
formation of dark matter halos in our model and compare the
results with observations. We assume matter domination and  we
take $T_\chi < m_0^\chi$, since we want the $\chi$ field to
represent dark matter. In that limit we can describe the $\chi$
field as a classical complex field. We consider a
spherically-symmetric metric

\begin{equation}
ds^2 = e^{2 u} dt^2 - e^{2v} dr^2 - r^2 d\Omega^2
\end{equation}

It is shown in ref.\cite{Friedberg} that all stable field
configurations have the form

\begin{equation}
\chi = \frac{\sigma(r)}{\sqrt{2}} e^{i \omega t} .
\end{equation}
Moreover one can take the Newtonian limit for the gravitational
field taking $u \simeq - v \simeq \Phi$. Under these assumptions,
the evolution of the gravitational potential is given \cite{Arbey}
by

\begin{equation}
\triangle \Phi = 4 \pi G \left(\rho^\chi_{eff} + \rho_b \right)
\end{equation}
where $\rho_b$ is the baryon energy density and

\begin{equation}
\rho_{eff}^\chi = 2 \omega^2 \sigma^2 - \frac{1}{2} m_0^{\chi 2}
\sigma^2 - \frac{1}{2} h \sigma^4
\end{equation}
The equation for the radial function $\sigma$ is \cite{Arbey}

\begin{equation}
\triangle \sigma + \left( 1- 4 \Phi  \right) \omega^2 \sigma -
\left( 1- 2 \Phi  \right) \left( \frac{m_0^{\chi 2}}{2} \sigma + h
\sigma^3 \right) = 0
\end{equation}

In the limit $\Lambda \equiv h/\left(4 \pi \, G \, m_0^{\chi 2}
\right) \gg 1 $ and neglecting  baryon contribution, one has the
approximate solution for $r/L_H \leq \pi$ \cite{Arbey}

\begin{equation}\label{solutionsigma}
\begin{array}{ll}
\sigma(r) = \sigma_0 \sqrt{ \frac{sin(r/L_H)}{(r/L_H)}}\\
\\
r \phi'(r) = v(r) = 2\pi \Lambda \frac{\sigma_0^2}{M_p^2} \left[
\frac{sin(r/L_H)}{(r/L_H)} - cos(r/L_H)        \right]\\
\\
L_H \equiv h^{1/2}\, \frac{M_p}{m_0^{\chi 2}}
\end{array}
\end{equation}
where $M_p$ is the reduced Planck mass. The solution
(\ref{solutionsigma}) is valid with the requirement that
$\Lambda^{-1} \ll \sigma_0/M_p \ll \Lambda^{-1/2}$. The mass of
the $\chi$ particles can be expressed as a function of $L_H$ and
of the coupling constant $h$ as

\begin{equation}\label{m0arbey}
m_0^\chi = h^{1/4} \, \sqrt{\frac{M_p}{L_{H}}} = 3 \, h^{1/4} \,
\sqrt{\frac{Kpc}{L_{H}}} \, eV
\end{equation}

The approximate solution (\ref{solutionsigma}) is valid for $r/L_H
\leq \pi$ where one has $\rho^\chi(r)/\rho^{\chi}(0) =
\sigma^2/\sigma_0^2 \sim sin(r/L_H)/(r/L_H)$ \cite{Arbey}. We can
say nothing about $\rho^\chi(r)$ for $r/L_H > \pi$ and we cannot
exclude that the exact solution can give a value of
$\rho^\chi(r)/\rho^\chi(0) \sim 1$ at some $r \gg L_H$. Therefore
the size of the halo could be many orders of magnitude greater
than $L_H$. That means that at this level of analysis, $L_H$ gives
a lower limit for the dark matter halo sizes. Therefore any value
of $L_H \leq 100 \, Kpc$ is acceptable, since there is no evidence
of halos of size much less than $100 \, Kpc$. In the next section
we will show that $\Delta^{eff}_\nu$ is very sensitive to the
choice of $L_{H}$ and that any value of $L_H \leq 0.1 \, Kpc$
gives a $\Delta_\nu^{eff}$ compatible with BBN bounds.

\section{Realistic model}\label{sectionrealisticmodel}

Let us show  how it is possible to construct a realistic model for
dark matter making use of the picture described in precedent
sections. Since we want to study the dependence of $\Delta_\nu^c$
and $\Delta_\nu^{th}$ on $L_H$, we parameterize $L_H$ and
$m_0^\chi$ as

\begin{equation}\label{m0nh}
L_{H} \simeq \frac{0.1}{\, n^2} \, Kpc, \qquad \qquad m_0^\chi
\simeq 10\,n \, h^{1/4}
\end{equation}

where $n$ will be fixed later. First we constrain the temperature
of the thermalized $\chi$ particles with the BBN bounds on
$\Delta^{eff}_\nu$. Since we expect that $m_0^\chi \ll
T_\chi^{BBN}\simeq k \, T_{\gamma}^{BBN}$, where $T_\gamma^{BBN}
\simeq 0.1-10 \, MeV$ is the photons temperature at BBN, therefore
$\rho^\chi_{th}$ will evolve as radiation at BBN and then it will
contribute to the effective number of extra neutrinos. Imposing
the condition $\Delta_\nu^{th} \leq 1$ \cite{BBN}, from
Eq.(\ref{extrarelativisticNthermal}) one obtains

\begin{equation}\label{boundk}
\frac{T_\chi}{T_\gamma} \equiv k \leq 0.8
\end{equation}

Note that as long as $T_\chi >  m_0^\chi$ one has $T_\chi \sim
1/a$ and $k$ maintains constant. Moreover from Eq.(\ref{boundk})
one has that $\rho^\chi_{th}/\rho_{rel} \leq k^4/g_r \leq 0.3
\cdot k^4$, were $g_r \geq 3.36$ is the relativistic degree of
freedom and $\rho_{rel}$ is the energy density of relativistic
particles. Now we should impose that at radiation-matter  equality
the condensate evolves as matter, and this implies that

\begin{equation}\label{t1condition}
T^{eq}_\chi = k \, T^{eq}_\gamma \leq T_{1\chi}
\end{equation}
where $T_\gamma^{eq}\simeq 0.69 \, eV$ is the temperature of
photons at radiation-mater equality. Moreover, as consistence
condition for the model, one has to require that
\begin{equation}\label{t2condition}
T_{1\chi} \gg T_{2\chi}
\end{equation}
We will check the validity of (\ref{t1condition}) and
(\ref{t2condition}) later on.

We want to stress two important facts. First the condition for the
condensate to evolve as radiation, i.e. $\rho^\chi_{c} \sim 1/a^4$
is $T_\chi \gg T_{1\chi} \equiv m_0^\chi/\alpha \gg m_0^\chi$,
therefore the condensate can evolve as matter at temperatures well
above $m_0^\chi$. Second the thermalized $\chi$ particles do not
have the same temperature of radiation but $T_\chi/T_\gamma = k
\leq 0.8$. This implies that at radiation-matter equality one has
$T_\chi^{eq} = k \, T_\gamma^{eq} \leq 0.56 \, eV$. Since we want
the condensate to represent dark matter, we impose that at
radiation matter equality $\rho_\chi^{c\, eq} = m_0^\chi \,
Q^{eq}_c= \rho_{DM}^{eq} \simeq 0.323 \, eV^4$. Since
$Q_c/T_\chi^3$ is constant for $T_\chi \gg T_{2\chi} \equiv
m_0^\chi$ and using Eq.(\ref{m0nh}) one has

\begin{equation}\label{Q-T espression}
\frac{Q_c}{T_\chi^3} = \frac{Q^{eq}_c}{T_\chi^{eq3}} = \frac{ 9.5
\times 10^{-2}}{n \, k^3 h^{1/4}}
\end{equation}

Assuming that $Q_c/T_\chi^3 \geq 1$ (since $h \ll 1$) and using
Eq.(\ref{m0nh}) again, one also has

\begin{equation}\label{alpha expression}
\alpha \simeq \frac{0.57 \, h^{1/4}}{n^{1/3} k}
\end{equation}

and from Eq.(\ref{extrarelativisticNcondensate}) one has

\begin{equation}
\Delta_\nu^c \simeq \frac{0.72}{n^{4/3}}
\end{equation}

Note that $\Delta_\nu^c $ depends only on $n$ and it is
independent of the other parameters of the model. This implies
that $\Delta_\nu^c$ is determined only from the choice of $L_H$.
Therefore one can choose $L_H$ in such a way that it gives a value
of $\Delta_\nu^c$ in the BBN bound. Imposing $\Delta_\nu^c \leq 1$
\cite{BBN} one obtains the  $n \geq 0.78$. We should cheek the
conditions given in Eqs.(\ref{t1condition}) and
(\ref{t2condition}). By use of  the expressions

\begin{equation}\label{temperatures}
\begin{array}{ll}
T_{1\chi} \simeq 17.6 \, k \, n^{4/3} \, eV \\
\\T_{2\chi} \simeq 10 \, n \, h^{1/4} \, eV\\
\\
T_\chi^{eq} = 0.69 \, k \, eV
\end{array}\end{equation}

one can check that Eq.(\ref{t1condition}) implies that $n \geq
0.088$ and Eq.(\ref{t2condition}) implies that $h \ll 10 \, k^4 \,
n^{4/3}$.

Let us take $n \simeq 0.8$ and $k \simeq 0.3$ in what follows.
With such a value of $n$ one obtains $L_H \simeq 0.17 \, Kpc$ that
is well below the typical size for dark matter halos and therefore
it is compatible with astrophysical observations. From
Eq.(\ref{Q-T espression}) one has $Q_c/T_\chi^3 \sim 10 \,
h^{-1/4} \geq 1$ for any $h <1$, therefore Eq.(\ref{alpha
expression}) is correct. Moreover one has $R \simeq h^{-1/4} \geq
0.2 \, h^{1/2}$, so Eq.(\ref{Rcondition2}) is fulfilled and the
values of $k \simeq 0.3$ and $n \simeq 0.8$ are compatible with
the condensate formation at early times. We also obtain $\alpha
\simeq 2 \, h^{1/4}$, $m_0^\chi \simeq 8 \, h^{1/4} \, eV$ and
$\Delta_\nu^c \simeq 0.97$.  Since $n \gg 0.088$ the condition
(\ref{t1condition}) is fulfilled and Eq.(\ref{t1condition})
implies that  $h \ll 10^{-2}$ for $n \simeq 0.8$ and $k\simeq 0.3$
. Though it is not necessary, one can ask that at equality time
$\rho^\chi_{th}$ still evolves as radiation, i.e. $T_\chi^{eq} \gg
m_0^\chi$, obtaining $h \ll 10^{-7}$. The values of $T_\chi$ and
$T_\gamma$ at $t_1$, $t_2$ and at matter-radiation equality are
resumed in table \ref{table1}. In table \ref{table2} we show the
values of $\alpha$, $m_0^\chi$ and $T_{\chi 2}$ for different
values of $h$. We stress that $m_0^\chi \sim 1-10^{-2} \, eV$ for
$h \sim 10^{-4}-10^{-12}$, though in the case of a scalar field
with no self-interaction, one needs an extremely light mass
$m^\chi_0 \sim 10^{-22} \, eV$  to avoid the formation of dark
matter halos of an excessively small size.

As we have already stressed,  both the values of $\Delta_\nu^c$
and $L_H$ only depends on $n$ as

\begin{equation}\label{delta e lh}
\Delta_\nu^c \simeq 3.34 \, \left( \frac{L_H}{Mpc}  \right)^{2/3},
\qquad L_H \simeq \frac{0.1}{n^2} Mpc
\end{equation}
and from Eq.(\ref{delta e lh}) it is evident how it is possible to
lower the value of $\Delta_\nu^c$ diminishing $L_H$. This means
that any value of $n \geq 0.8$ will give a $\Delta_\nu^{eff}$ in
the BBN bounds and a value of $L_H \leq 0.16 \, Kpc$ well below
the typical size of dark matter halos.

We note that in \cite{Arbey} the authors  take $L_H$ of the order
of the core of dark matter halos, i.e. $L_H \simeq 10 \, Kpc$.
They also take the coupling in the interval $h \simeq 1-10^{-4}$,
obtaining a mass $m_0^\chi \sim 1 \, eV$ and they show that such
values of the mass and coupling give a number of effective extra
neutrinos $\Delta^{eff}_\nu $ that exceeds the BBN bound. This
result is in agreement with our analysis but we have shown that it
is possible to take smaller  $L_H \leq 0.1 \, Kpc$ to lower the
value of $\Delta_\nu^{eff}$ below BBN bounds. Of course this is
possible since, as we have discussed in section \ref{dmhalos},
$L_H$ is a lower bound for the typical size of dark matter halos
and therefore any $L_H \leq 100 \, Kpc$ is in agreement with
astrophysical observations.

We stress that we have used the approximate (and incomplete)
solution given in Eq.(\ref{solutionsigma}) to have a lower limit
for the dark matter halo sizes. Of course this analysis is
incomplete in many respects, since it does not take into account
the formation of dark matter halos as evolving from linear
perturbations nor how the cosmological evolution of the universe
influences this process. Moreover the presence  of a residual
$\rho^\chi_{th}$ as well as a cosmological constant, baryons and
radiation, were not considered. Therefore we think that an
analysis of  dark matter halos formation that takes into account
all of these considerations will be very useful to further
constrain the model.

In particular, we note that the conclusions  of this section are
based on the relation given in Eq.(\ref{m0arbey}) between
$m^\chi_0$ and $L_H$. An analysis of dark matter halos formation
different from that described in section \ref{dmhalos}, can
considerably change Eq.(\ref{m0arbey}) and therefore it can give
less stringent constraints on the parameters of our model.

\begin{table}[ht]
\begin{tabular}{|l|l|l|l|}
\hline
t & $t_1$ & $t_2$ & $t_{eq}$ \\
\hline
$T_\chi$ & $0.39 \, eV$ & $8 \, h^{1/4} \, eV$ & $0.21 \, eV$\\
\hline
$T_\gamma$ & $13 \, eV$ & $26.7 \, h^{1/4} \, eV$ & $0.698 \, eV $\\
\hline
\end{tabular}
\caption{Values of $T_\chi$ and $T_\gamma$ at three different
times: $t_1$ when $T_\chi = T_{1\chi}$, $t_2$ when $T_\chi =
T_{2\chi}$ and $t_{eq}$ at radiation-matter equality.}
\label{table1}
\end{table}

\begin{table}[ht]
\begin{tabular}{|l|l|l|l|}
\hline
h & $10^{-4}$ & $10^{-8}$ & $10^{-12}$ \\
\hline
$\alpha$ & $0.2$ & $ 0.02$ & $0.002$\\
\hline
$m_0^\chi$ & $0.8 \, eV$ & $0.08 \, eV $ & $0.008 \, eV$\\
\hline
$T_{2\gamma}$ & $2.67 \, eV$ & $0.267 \, eV $ & $0.002 \, eV$\\
\hline
\end{tabular}
\caption{Values of $\alpha$, $m_0^\chi$ and $T_{2\gamma}$ for
different values of the coupling $h$.} \label{table2}
\end{table}

\section{Conclusions}\label{sectionconclusions}

We have shown how it is possible to use a complex scalar field
with self-interactions $h\, |\chi|^4$ in order to obtain a
realistic model for dark matter. In this model, dark matter is
described as a condensate of $\chi$ particles that forms just
after reheating and dominates at late times.

As  pointed out, the presence of the self-interaction is very
important in the model. First, it is essential to explain the
condensate formation. In fact the $\chi$ field is produced at
reheating with a charge asymmetry, and under the conditions
discussed in section \ref{sectioncondensate}, self-interactions
drive the formation of a $\chi$-particle condensate. Therefore,
just after reheating the $\chi$ field configuration is that of a
$\chi$-particle condensate in equilibrium with a thermalized gas
of $\chi$ and $\bar\chi$ particles.

Second, due to self-interactions, the $\chi$ mass has thermal
corrections that are important at high temperatures $T_\chi \gg
T_{1\chi} \gg m_0^\chi$, where $m^\chi \simeq m_{th}^\chi \sim
T_\chi$. This implies that as long as $T_\chi \gg T_{1\chi}$ one
has $\rho^\chi_c \sim T_\chi^4$ and the $\chi$ field behaves as
radiation. This  explains why dark matter dominates at late times.
In fact, one can choose the parameters of the model properly in
order to ensure that the $\chi$ condensate begins to evolve as
matter with $\rho^\chi_c \sim T_\chi^3$ just before radiation
matter equality and that at equality one has the right value
$\rho^\chi = \rho_{DM}\simeq 0.323 \, eV^4$. We have also given a
lower bound $L_H$ for the size of dark matter halos and we have
studied the dependence of the contribution of the $\chi$ field to
the effective number of extra neutrinos $\Delta_\nu^{eff}$ on
$L_H$. We have shown that, according to Eq.(\ref{delta e lh}), it
is possible to diminish $\Delta_\nu^{eff}$ just lowering the value
of $L_H$ and that any $L_H \leq 0.1 \, Kpc$ gives a
$\Delta_\nu^{eff}$ within BBN bounds. Since $L_H$ is a lower bound
for dark matter halo  sizes, any value $L_H \leq 100  \, Kpc$ is
acceptable. In section \ref{sectionrealisticmodel} we have
constructed a realistic model, choosing $L_H \simeq 0.1 \, Kpc$,
$T_\chi/T\gamma \simeq 0.3$, and the coupling in the interval $h
\simeq 10^{-4}-10^{-12}$. With such values of the parameters the
condensate begins to evolve as matter with $\rho^\chi_c \sim
T_\chi^3$ at $T_\gamma \simeq 13 \, eV$ and it gives the right
value $\rho^\chi_c = \rho_{DM} \simeq 0.323 \, eV^4$ at equality
epoch $T_\gamma \simeq 0.698 \, eV$.

Therefore, we conclude that, at the present level of analysis, our
model is in agreement with cosmological and astrophysical
observations. Of course a more profound analysis of the evolution
and growth of cosmological perturbations in that model is still
missing, but we think that such a study would be very useful to
further constrain the model with cosmological data.

{\bf Acknowledgements}: I would like to thank M. Lattanzi for
useful discussions at the very beginning of this work. I would
like to thank T. Matos, A. Melchiorri, G. Amelino-Camelia, S.D.
Odintsov and M. de Llano for useful discussions during the writing
of this Letter.

\end{document}